# Discovering and Characterizing the Planetary Systems of Nearby Stars
*The scientific need for medium aperture space coronagraph observations*

Authors: T. Greene[1], K. Cahoy[1], O. Guyon[2], J. Kasting[3], M. Marley[1], M. Meyer[2], S. Ridgway[4], G. Schneider[2], W. Traub[5], N. Woolf[2]

Affiliations:
1. NASA – Ames Research Center
2. University of Arizona
3. Penn State University
4. National Optical Astronomy Observatory
5. Jet Propulsion Laboratory

Lead author contact:     Tom Greene
                         NASA Ames Research Center
                         tom.greene@nasa.gov
                         650-604-5520

# 1. Exoplanet Opportunities and Context

The study of extrasolar planets has exploded in the first decade of the 21$^{st}$ Century. There are now over 330 known exoplanets, nearly all with masses constrained by radial velocity (RV) measurements. Many of these exoplanets have projected masses (m sin i) about that of Jupiter, but over a dozen have projected masses less than Neptune (below 20 Earth masses). Recent improvements in RV precision have revealed an increasing number of very low mass planets, including potential Super-Earths, and many more will likely be discovered in the near future. The Kepler mission will soon be launched, and it should determine the frequency of planets as small as the Earth in habitable zones of stars in our region of the Galaxy.

Exoplanet characterization is also advancing at least as quickly as exoplanet discovery. Over 50 planets have been found to transit the disks of their host stars. When combined with RV measurements, photometric transit observations have revealed exoplanet masses, radii, densities, and orbital periods. This information alone has constrained the structures of these planets, finding that several have "inflated" radii that are larger than expected from model predictions. HST transmission spectroscopy (Charbonneau et al. 2002, Barman 2007, Swain et al. 2008b) and Spitzer secondary eclipse observations (e.g., Knutson et al. 2007) are revealing the temperatures, atmospheric temperature profiles, surface winds, and compositions of several exoplanets. Indeed, we are entering a new era of exoplanetary science.

This rapid progress is very exciting, and much more should be learned about exoplanets over the next decade. Precision RV observations with HARPS, HARPS-N, and new facilities (e.g., APF, N-EDI / T-EDI) will likely result in discovery of many more Super-Earth mass planets around nearby stars. More sensitive transit surveys are underway, in development, or under study, and some may reveal whether there are small-to-large planets in the habitable zones of nearby M stars. The James Webb Space Telescope (JWST) should be able to obtain very high quality spectra of transiting Jupiter-size planets over near- to mid-infrared (IR) wavelengths and should be able to detect earth-sized planets transiting M dwarf stars. The MMT and Keck nulling interferometers will also be able to detect the mid-IR thermal emissions of circumstellar dust disks with masses down to 10 – 100 times that of our own (10 – 100 zodis, respectively).

We are now on the verge of detecting and characterizing Earth-analog and Super-Earth rocky planets in the habitable zones around nearby stars, an age-old quest of astrophysics and mankind. We could begin a modest aperture space mission in the next decade that would address the next set of fundamental exoplanetary science questions with direct imaging observations. What are the temperatures and compositions of the closest Neptune-to-Jovian mass planets discovered by radial velocities that are not excessively irradiated by their host stars? Which of the closest stars in our immediate solar neighborhood have Earth-like or Super-Earth planets in habitable zones? Are the low resolution spectra of these small planets similar to the terrestrial planets in our own solar system, or are they more bizarre water or ice worlds? How much exozodiacal dust is there around nearby stars, and how symmetrically is it distributed? What is the composition of exozodiacal dust, and what can we learn about unseen planets from its distribution?

Observations that address these issues will also shed much light on the formation and evolution of stars and planetary systems including our own Solar System. The simultaneous observation of planets and dust disks should reveal much about planetary architectures, planetary compositions, and the creation and dispersal of debris disks. It is likely that suitable observations of the planetary systems around the nearest stars will drive the development of planetary systems theory (as has been the case for the past decade), and this will drive our understanding of these systems, their evolution, and their host stars.

## 2. Key Advances Required

The key advance needed to discover and characterize the planetary systems of nearby stars is new high spatial resolution, high contrast imaging data over the 400 – 800 nm visible band with modest spectral resolution, R ~ 15-20. These data can be obtained most efficiently by developing a modest aperture (~ 1.5-m) high performance space coronagraph mission. High contrast coronagraphic imaging observations (rejecting all but 1E-10 of the host star) with inner working angles (IWAs) of ~150 mas will be capable of probing down to the habitable zones of nearby FGK stars. Guyon et al. (2008) describe how utilizing a high efficiency Phase-Induced Amplitude Apodization (PIAA) coronagraph (e.g., Guyon 2003) in a moderate aperture (1.4-m) space telescope with several wavefront controlled channels can provide the needed imaging data. The high efficiency, high contrast, and small IWA of the PIAA coronagraph are essential for making adequate observations with such a small (and inexpensive) space telescope.

A moderate-sized space telescope with a high performance coronagraph would provide a capability threshold that is adequate for addressing numerous important questions about the planetary systems of nearby stars. Herein we consider the capabilities of a 1.4-m aperture mission. Such an observatory would have adequate sensitivity to detect Earth- and Super-Earth size (1 and 2 Earth radii) and Earth albedo planets around 7 – 20 (Earth – Super-Earth) nearby stars in the presence of 1 zodi disks in under 12 hours of integration time in a 100 nm wide V band filter. Even moderate signal-to-noise, low spectral resolution (R ~ 5 filters) photometric imaging data would be capable of distinguishing between the rocky, ice giant, and gas giant planets in our solar system (Fig. 1). Slightly higher spectral resolution (R~15) data would differentiate them much better (Fig. 2).

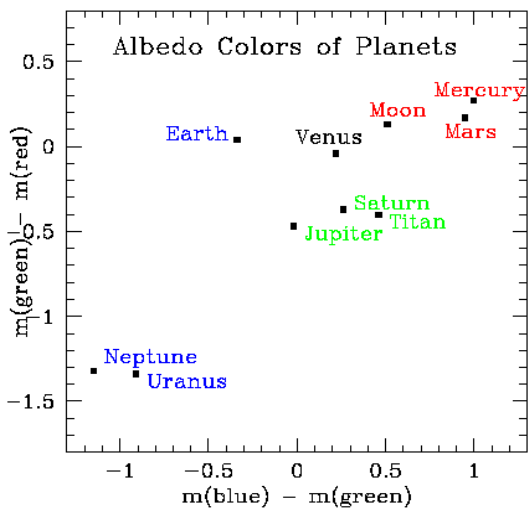

*Figure1: Colors of Solar System planets (courtesy of W. Traub)*

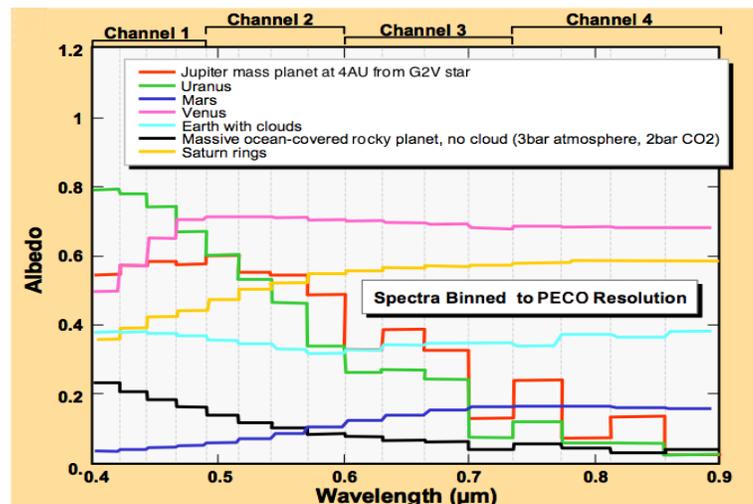

Figure 2: Planet albedos through R=15 spectral bands. Earth's atmosphere has a relatively constant albedo across the visible, with a slight absorption near 600 nm due to ozone. EGPs like Jupiter will have relatively flat spectra, with deep methane absorption in the red adjacent to bright continua arising from clouds. Cooler, lower gravity, and/or methane-rich ice giants like Uranus & Neptune are bluer and much darker in the red.

The spectral energy distributions of Super-Earth sized planets around the nearest ~10 stars could be characterized to moderate signal-to-noise (SNR ~ 20) in about a week each. Dust disks on the order of 1 zodi could be detected in hours, and their spatial extents and low resolution spectra could be measured in a few days. Jupiter analogs could be detected in under an hour for dozens of stars, and they could be characterized to SNR ~ 30 in under a week each. Numerous known RV planets could be detected and characterized; a simulated coronagraphic image of the planet 47 Uma b embedded in a 3 zodi disk is shown in Figure 3.

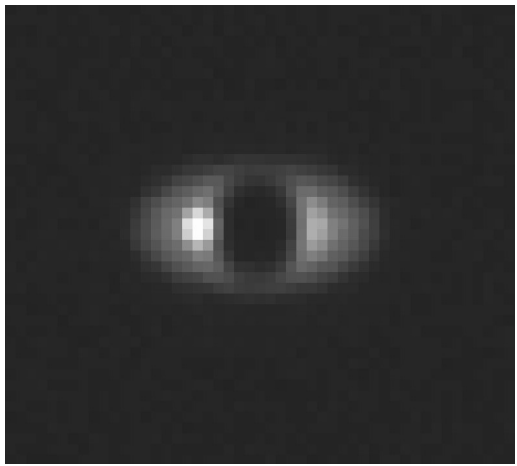

**Figure 3:** Simulation of 24 hr coronagraphic data showing the Jovian planet 47 Uma b (see Table 1) with 3 zodis of exododi dust in a disk with surface density $r^{-0.34}$ inclined 59 degrees. This is a simulation of $\lambda$= 550 nm light in a 100 nm bandpass with predicted PIAA performance of a 1.4-m aperture observatory with a PIAA coronagraph. 47 Uma is a G0V star at 14 pc distance. Photon noise and detector noise for an electron multiplying CCD have been added.

A moderate aperture coronagraph is likely the best way to address the scientific issues outlined in this white paper. It is true that JWST will be able to obtain high quality spectra of transiting gas giant planets (Greene et al. 2007), but it will be difficult for JWST to detect low mass dust disks (below hundreds of zodis) near habitable zones or study small planets. Transit observations are limited by the photon noise of the host star, and that is too great even for JWST to make much headway in studying small planets (Beckwith 2008). The recent Exoplanetary Task Force report is also being modified to reflect this fact. The signal to noise of photon noise limited observations increases as the square root of collecting area, so JWST should have intrinsic SNR of about 7 times greater than Spitzer. This should allow good quality spectra of gas giant planets that orbit their host stars closely, but it is not sufficient to characterize small planets or ones (e.g., solar system analogs) that do not transit their host stars.

## 3. Scientific Discovery Potential and Compelling Questions

The general area of the discovery and characterization of the planetary systems of nearby stars includes studying terrestrial planets in their habitable zones to giant planets and circumstellar debris disks. We now present 3 compelling questions in this field and examine how relatively modest coronagraphic imaging data (realistic for a mission started this next decade) might address them.

### *3.1. What are the numbers and properties of giant planets in a sample of nearby stars?*

The reflection spectra of mature giant planets are controlled by Rayleigh and Mie scattering from atmospheric gases, aerosols, and cloud particles, and by gaseous absorbers. Scattering of incident light usually dominates in the blue, giving way to absorption by the major molecular components at wavelengths greater than about 0.6 μm. The major absorbers in the optical are methane and, for warmer planets, water. Generally speaking, in strong molecular bands photons are absorbed before they can scatter back to space. In the continua between bands, photons scatter before they are absorbed. The continuum flux from a given object is thus controlled by Mie scattering from its clouds and hazes and Rayleigh scattering from the column of clear gas above the clouds. Figure 4 illustrates the significant impact that clouds can have on exoplanet spectra.

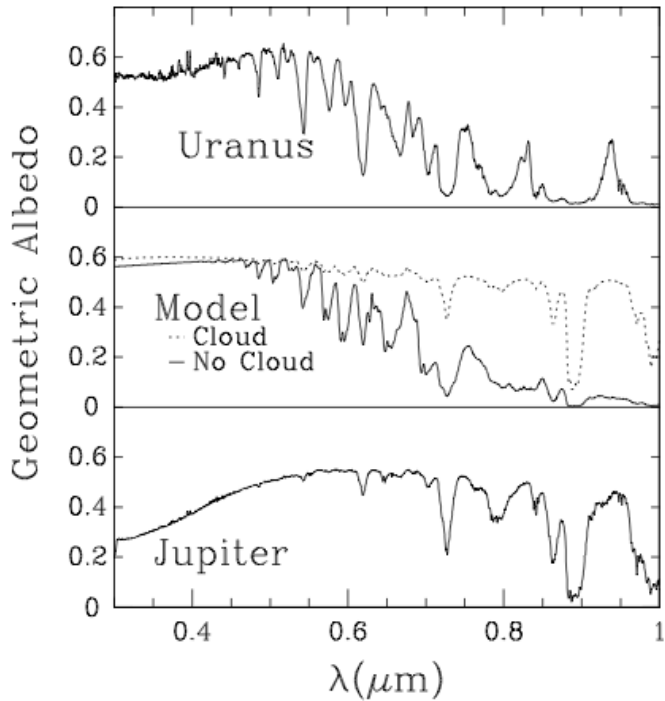

**Figure 4:** Geometric albedo spectra of Jupiter and Uranus (Karkoschka 1994), compared to model spectra for a Jupiter-mass, solar-composition planet with Jupiter's effective temperature (128 K). The model spectra demonstrate the importance of clouds in controlling the reflected spectra. Both models have a solar abundance of methane and no water. The cloudy model includes a stratospheric haze at 0.001 bar (with an optical depth $\tau = 0.1$) and a tropospheric cloud at 1 bar ($\tau = 5$). Marley et al (1999)

Thus visible-wavelength spectra of giant planets opens a door into their cloud structure (and by extension atmospheric temperature) and composition. For cold giants, like Jupiter and Saturn, the abundance of C will be constrained through the well-studied methane absorption features that dominate their optical spectra (Figure 4). In somewhat warmer atmospheres (younger or more massive planets or giants closer to their primaries or hotter primaries) ammonia will not condense but instead be present as a gas in the atmosphere. Such planets will sport water clouds and have a flat, bright continuum spectra punctuated by methane and ammonia absorption. Even warmer planets, without clouds, will show water vapor bands in the optical and be very dark in the red. Meanwhile "ice giants" (highly enriched atmospheres like Uranus and Neptune) will have their own unique spectral sequence characterized by very blue colors because of the overwhelming methane absorption in the red and should be easily distinguishable from their less enriched, more massive siblings (Figures 1 and 2). One challenge, however, is that composition information is found in the depth of absorption features, which are of course at less favorable contrast and lower S/N than the continuum. Thus composition determinations should focus on weak to moderate strength bands (e.g., at 0.62 μm or 0.54 μm) rather than the strongest bands (e.g., the 0.89 μm methane band) which will be quite dark.

Color will provide the zeroth order characterization of imaged exoplanets. A set of moderate and wide band filters spanning the 400 – 800 nm wavelength region should also be used to measure the strong absorptions of NH3, CH4, and H2O to constrain giant planet temperatures and compositions.

Low resolution spectra alone are not sufficient to characterize EGPs; we must also know the mass of a planet to determine its temperature and composition with high precision. A moderate aperture space PIAA coronagraph will be able to observe over a dozen known radial velocity planets with known m *sin i* masses (Table 1). These planets span the 0.5 – 1 Mjup projected mass

*Astro2010 Science White Paper: Discovering Planetary Systems of Nearby Stars* 4

range at distances 1.7 – 6 AU from their host stars. Two well-separated observations of each will constrain their inclinations and thus their masses. These systems should form a good set of planets that can be used to interpret observations of other EGPs with less well known masses.

Table1: Known RV exoplanets easily observable with a moderate aperture coronagraph

| Planet Name | Pl. Mass | Period (d) | a (AU) | sep`` | 550nm/D | Dist (pc) | St. Sp T | M* | St. Mag. V. | Pl mag V | Contrast |
|---|---|---|---|---|---|---|---|---|---|---|---|
| Epsilon Eridani b | 1.55 | 2502 | 3.39 | 1.06 | 13.07 | 3.2 | K2 V | 0.8 | 3.73 | 25.7 | 1.6E-09 |
| 55 Cnc d | 3.84 | 5218 | 5.77 | 0.43 | 5.31 | 13.4 | G8 V | 1.0 | 5.95 | 29.1 | 5.5E-10 |
| HD 160691 c | 3.1 | 2986 | 4.17 | 0.27 | 3.36 | 15.3 | G3 IV-V | 1.1 | 5.15 | 27.6 | 1.1E-09 |
| Gj 849 b | 0.82 | 1890 | 2.35 | 0.27 | 3.3 | 8.8 | M3.5 | 0.4 | 10.42 | 31.6 | 3.3E-09 |
| HD 190360 b | 1.5 | 2891 | 3.92 | 0.25 | 3.04 | 15.9 | G6 IV | 1.0 | 5.71 | 28.0 | 1.2E-09 |
| 47 Uma c | 0.46 | 2190 | 3.39 | 0.24 | 2.99 | 14.0 | G0V | 1.0 | 5.1 | 27.1 | 1.6E-09 |
| HD 154345 b | 0.95 | 3340 | 4.19 | 0.23 | 2.86 | 18.1 | G8V | 0.9 | 6.74 | 29.2 | 1.0E-09 |
| Ups And d | 3.95 | 1275 | 2.51 | 0.19 | 2.3 | 13.5 | F8 V | 1.3 | 4.09 | 25.4 | 2.9E-09 |
| Gamma Cephei b | 1.6 | 903 | 2.04 | 0.17 | 2.14 | 11.8 | K2 V | 1.4 | 3.22 | 24.1 | 4.4E-09 |
| HD 62509 b | 2.9 | 590 | 1.69 | 0.16 | 2.02 | 10.3 | K0IIIb | 1.9 | 1.15 | 21.6 | 6.4E-09 |
| HD 39091 b | 10.35 | 2064 | 3.29 | 0.16 | 1.97 | 20.6 | G1 IV | 1.1 | 5.67 | 27.6 | 1.7E-09 |
| 14 Her b | 4.64 | 1773 | 2.77 | 0.15 | 1.89 | 18.1 | K0 V | 0.9 | 6.67 | 28.2 | 2.4E-09 |
| 47 Uma b | 2.6 | 1083 | 2.11 | 0.15 | 1.86 | 14.0 | G0V | 1.0 | 5.1 | 26.1 | 4.1E-09 |

We have computed that Jupiter analogs (same albedo, 5 AU orbits) could be detected with 20% probability (random positions in their orbits) around over 200 nearby stars in under 6 hours of integration time each (100 with integration times under an hour). Thus a survey of 200 stars would yield on the order of 40 giant planets (for eta_planet = 1) and could be completed in under a year, including time for spectral characterization of detected planets.

### 3.2 What is the frequency, masses, distribution, and composition of circumstellar dust disks around nearby stars?

Disks of circumstellar material are both the progenitors and outcomes, of the processes of planet formation and planetary system evolution. Hundreds of sun-like stars have been found to exhibit excess IR emission attributable to dusty circumstellar debris. The vast majority of these systems, however, remain spatially unresolved (Meyer et al. 2008). Dust temperatures and covering fractions may be estimated from long wavelength spectral energy distributions (SEDs), but the *inferred* locations of the thermally emissive orbiting dust grains depend upon the properties assumed for the particles. Reasonable ranges of particle sizes and compositions result in models for the dust-producing planetesimals that vary by an order of magnitude in orbital radius and total dust mass (e.g., Fig. 5 in Hines et al. 2006). High-resolution scattered-light images of debris disks will reveal the morphology of the disks and trace the location of the dust-producing planetesimals, but such images today are rare due to the lack of suitable high-contrast, high resolution, small IWA imaging systems.

Currently deployed "high contrast" imaging technologies can detect only the largest, most massive and brightest circumstellar disks, and cannot effectively probe their innermost regions. The few resolved images obtained to date have provided crucial insights into the formation, evolution, and architectures of exoplanetary systems; but they are just the tip of the iceberg waiting to be fully revealed. The existing sparse sample of debris disks imaged with scattered starlight represents only the youngest (10–100 Myr), or extremely anomalous, disk systems around older stars. Observations with a new moderate aperture space coronagraph are required to address key questions :

• What is the amount and distribution of circumstellar dust around the stars in the solar neighborhood? How is dust distributed in their habitable zones and how does this impact direct planet detection (also relevant for future larger missions)?

• Are there dynamical structures visible in the circumstellar disks of nearby stars, and what can we learn about unseen planets from zones of different material or disk gaps?

• What are the physical properties of the exozodiacal debris material; what are the grain properties and distributions?

Starlight-suppression with a 1.4-m aperture PIAA coronagraph will achieve disk-imaging contrasts of ~$10^{-10}$ exterior to a 130 mas IWA (at 0.4 microns), an increase of ~ 5 to 6 orders of magnitude in contrast beyond what *HST* and ground-based 6–10 m telescopes with current adaptive optics systems can provide. *JWST* will have contrast-limited performance no better than *HST*. This IWA and spatial resolution is similar to what *the Large Binocular Telescope Interferometer (LBT-I)* will provide at 11 microns, enabling a comparison of circumstellar dust in scattered light and thermal IR emission at unprecedented spatial resolution. This moderate aperture space coronagraph will also be at least 10 times as sensitive to interstellar dust in habitable zones as LBT-I. This high sensitivity to light-scattering circumstellar debris at this small IWA will provide strong constraints on systemic dust mass-loss rates, and observationally test models of dust production throughout the epochs of planet formation, and their subsequent dynamical evolution.

Such an observatory will be capable of providing the first direct images in scattered light of debris disks (the analogs of our solar system's zodiacal dust cloud, asteroid and Kuiper belts) around a large sample of nearby FGK stars (~200, same as for the gas giant survey). The disk sizes will likely vary from unresolved (interior to the IWA) to much more than 10 arc-seconds from some stars (e.g., the very large disk circumscribing Fomalhaut).

### *3.3 What planets are in the habitable zones of the closest stars, down to the size of Earths?*

A 1.4-m PIAA coronagraph will be capable of detecting even small planets (1 – 2 $R_\oplus$) with Earth albedos in the habitable zones of about two dozen of the nearest stars. This is a large enough sample to probe several significant questions:

• Which of the closest stars have Earth-like planets in or near their habitable zones?

• What are the broad-band colors of small planets; are they similar to ones in our own solar system?

• Are any very strong spectral features are seen in the atmospheres of these planets; do any show H2O, O2, or any other molecules needed for life as we know it?

The atmospheres of such warm terrestrial exoplanets may show spectral features produced by water vapor. These can be seen in the albedo spectrum of Earthshine (Woolf et al. 2002). The principal bands in the visible / near-IR spectrum are H2O at 825 +/- 20nm and 720 +/- 20nm.

Spectral features produced by molecules containing oxygen are also discernable at low spectral resolution. The most prominent at low spectral resolution is the Chappuis band of ozone at 590 +/- 50nm. Next most prominent is the very deep oxygen A band at 760 +/- 10nm. Detecting O3, H2O, and O2 in the atmospheres of terrestrial planets will be possible for those orbiting the nearest stars. The H2O and O3 features are broad and shallow (~10% deep), requiring SNR ~ 30 on the continuum for detection.

Our simulations show that any small planets in or near habitable zones of 20 of the nearest stars would have a 20% chance of detection in 6 – 12 hours of integration time with a moderate aperture space coronagraph. Thus there would be a 90% chance of detecting each one in a total of 10 uncorrelated visits. Therefore a complete survey and repeat followup characterization could be completed in about a year of real time. These observations will likely include short term monitoring for variation with rotation and longer term monitoring for seasonal effects (perhaps snow), phase effects in atmospheric scattering, and constraining orbits.

## Summary

Significant advances in the discovery and characterization of the planetary systems of nearby stars can be accomplished with a moderate aperture high performance coronagraphic space mission that could be started in the next decade. Its observations would make significant progress in studying terrestrial planets in their habitable zones to giant planets and circumstellar debris disks, also informing the design of a more capable future mission. It is quite exciting that such fundamental exoplanet science can be done with relatively modest capabilities.